\documentclass[%
reprint,
superscriptaddress,
 amsmath,amssymb,
 aps,
prl,
]{revtex4-1}

\usepackage{graphicx}
\usepackage{dcolumn}
\usepackage{bm}
\usepackage[mathlines]{lineno}

\usepackage[utf8]{inputenc}
\usepackage[T1]{fontenc}
\usepackage{mathptmx}
\usepackage{xcolor}

\begin{document}


\title{From Weak Antilocalization to Kondo Scattering in a Magnetic Complex Oxide Interface}

\author{Xinxin Cai}
\affiliation{School of Physics and Astronomy, University of Minnesota, MN 55455, USA}
\author{Jin Yue}
\affiliation{Department of Chemical Engineering and Materials Science, University of Minnesota, MN 55455, USA}
\author{Peng Xu}
\affiliation{Department of Chemical Engineering and Materials Science, University of Minnesota, MN 55455, USA}
\author{Bharat Jalan}
\affiliation{Department of Chemical Engineering and Materials Science, University of Minnesota, MN 55455, USA}
\author{Vlad S. Pribiag}
\email{vpribiag@umn.edu}
\affiliation{School of Physics and Astronomy, University of Minnesota, MN 55455, USA}

\date{\today}

\begin{abstract}
Quantum corrections to electrical resistance can serve as sensitive probes of the magnetic landscape of a material. For example, interference between time-reversed electron paths gives rise to weak localization effects, which can provide information about the coupling between spins and orbital motion, while the Kondo effect is sensitive to the presence of spin impurities. Here we use low-temperature magnetotransport measurements to reveal a transition from weak antilocalization (WAL) to Kondo scattering in the quasi-two-dimensional electron gas formed at the interface between SrTiO$_3$ and the Mott insulator NdTiO$_3$. This transition occurs as the thickness of the NdTiO$_3$ layer is increased. Analysis of the Kondo scattering and WAL points to the presence of atomic-scale magnetic impurities coexisting with extended magnetic regions that affect transport via a strong magnetic exchange interaction. This leads to distinct magnetoresistance behaviors that can serve as a sensitive probe of magnetic properties in two dimensions.      
\end{abstract}

\maketitle

\begin{figure}
\includegraphics{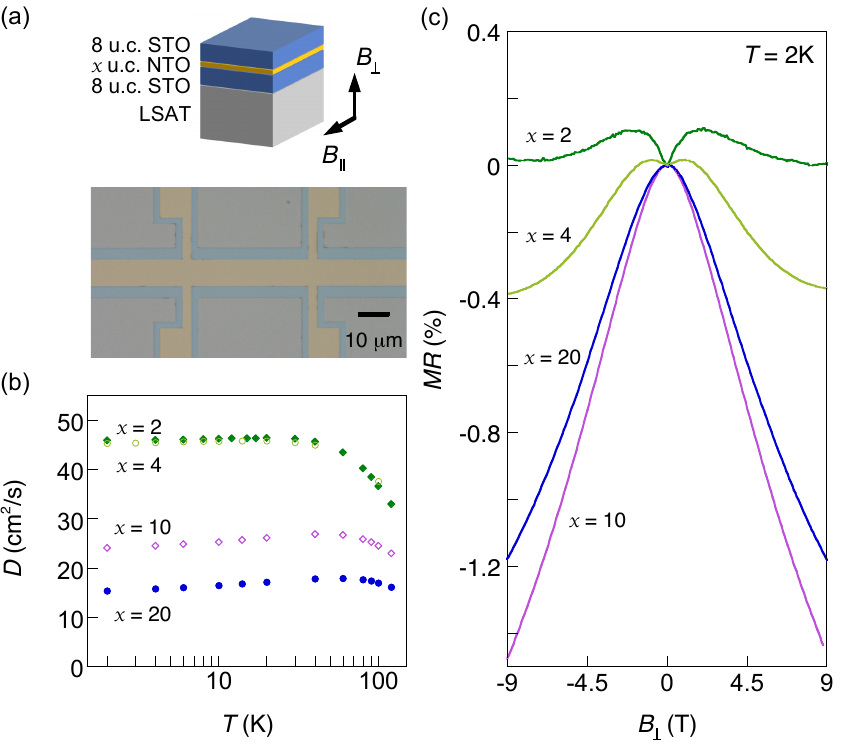}
\caption{\label{Fig1} (a) Top: A schematic of the capped STO(8 u.c.)/NTO($x$ u.c.)/STO(8 u.c.)/LSAT(001) heterostructure, where $x=$2, 4, 10, 20. The directions ofapplied fields are indicated with respect to the heterointerface. Bottom:
False-color optical image of a typical Hall-bar sample prepared on
the heterostructure. The etched regions are indicated in blue. (b) The
temperature dependence of the diffusion coefficient $D$. (c) Magnetoresistance as a function of the perpendicular magnetic field $B_{\bot}$ measured at $T=2$~K. }
\end{figure}

Conducting interfaces between SrTiO$_3$ (STO) and other complex oxides are an ideal system for investigating two-dimensional electron systems in the high-density regime\cite{ohtomo_high-mobility_2004,takizawa_photoemission_2006,moetakef_electrostatic_2011,xu_stoichiometry-driven_2014}. Our experimental system consists of epitaxial layers of SrTiO$_3$ and NdTiO$_3$ (NTO) \cite{xu_stoichiometry-driven_2014, xu_quasi_2016, xu_predictive_2016}, which host a high-density quasi-two-dimensional electron gas coupled to local ferromagnetic or superparamagnetic regions\cite{ayino_ferromagnetism_2018, cai_disentangling_2019}. The ferromagnetic order is thought to originate from spatially inhomogeneous canting of the antiferromagnetically-alligned spins in NTO\cite{sefat_anderson-mott_2006, ayino_ferromagnetism_2018}, which is a Mott-Hubbard insulator with a N$\acute{\text{e}}$el temperature of $\sim90$~K \cite{amow_structural_1996,sefat_effect_2006}. Here we focus on MBE-grown heterostructures with layer structure STO(8~u.c.)/NTO($x$~u.c.)/STO(8~u.c.)/(La,Sr)(Al,Ta)O$_3$ (LSAT)(001) (substrate), where $x=$2, 4, 10, 20 (Fig.~\ref{Fig1}(a)). The top STO layer protects the underlying heterostructure from degradation\cite{xu_predictive_2016}, however only the bottom NTO-on-STO interface is expected to contribute to transport, as STO-on-NTO interfaces are typically insulating\cite{xu_quasi_2016, ayino_ferromagnetism_2018}. In this system, itinerant electrons, which reside primarily on the STO side of the interface\cite{xu_quasi_2016}, experience the combined effect of $k$-cubic Rashba spin-orbit coupling and a magnetic exchange interaction of several 10's of Tesla due to the local, inhomogeneous magnetic order\cite{cai_disentangling_2019}. 

Previous work has focused on the case of thin NTO layers, where these effects give rise to a unique type of WAL correction to the conductance under the combined effect of the magnetic exchange and an applied magnetic field \cite{cai_disentangling_2019}. Here we report on a striking transition between this regime, where localization effects dominate the magnetotransport response, to a regime of predominant Kondo scattering, as the thickness of the NTO layer is increased. Interestingly, for the thickest NTO layers studied, the Kondo effect coexists with the interfacial exchange interaction, suggesting the presence in close proximity of both atomic scale spin impurities and larger magnetically-ordered regions.

The heterointerfaces are mesa-etched into $10\times20$ and $10\times40~\mu$m$^2$ Hall-bar devices by a combination of electron beam lithography and dry etching techniques. Fig.1(a) shows a typical device. To achieve ohmic contacts to the conducting interfaces, the cross-sections of the heterostructures are exposed by Ar ion milling and coated afterwards by Ti/Au electrodes via an angled deposition. The magnetotransport measurements are performed in a 9-T Quantum Design physical property measurement system (PPMS) at temperatures ($T$) down to 2~K. A rotational sample holder is used for applying magnetic fields at various angles with respect to the sample plane. Four-terminal resistance is measured using standard DC techniques with currents $\leq 0.5~\mu$A. 

Taking the effective mass of the conduction electrons to be $m^*=0.8 m_e$ \cite{mattheiss_effect_1972}, we estimate the diffusion constant $D$ for our samples with different NTO thicknesses based on the longitudinal resistance and Hall effect data. The estimated $D$ values are presented in Fig.~\ref{Fig1}(b) for a broad range of temperatures. 
Fig.~\ref{Fig1}(c) shows the sample magnetoresistance ($MR$) as a function of perpendicular field $B_{\bot}$ measured at $T=2$~K. The data for the 2~u.c.~sample has the characteristic shape of WAL, with sharp positive $MR$ clearly seen around zero field, and a maximum value at at larger field, around $B_\bot=2$~T. The WAL/WL contribution to the magetoresistance in $B_\bot$ can be analyzed to extract the dephasing and SOC parameters of the system, $B_\phi$ and $B_{so}$\cite{cai_disentangling_2019}. The focus of this work is on the dependence of the $MR$ on the NTO thickness. Fig.~\ref{Fig1}(c) shows four snapshots of this striking evolution. As is shown below, the pronounced negative $MR$ in the thicker samples originates from the interplay between  localization and the Kondo effect, with the $x=4$ sample revealing crossover behavior.

\begin{figure} 
\includegraphics{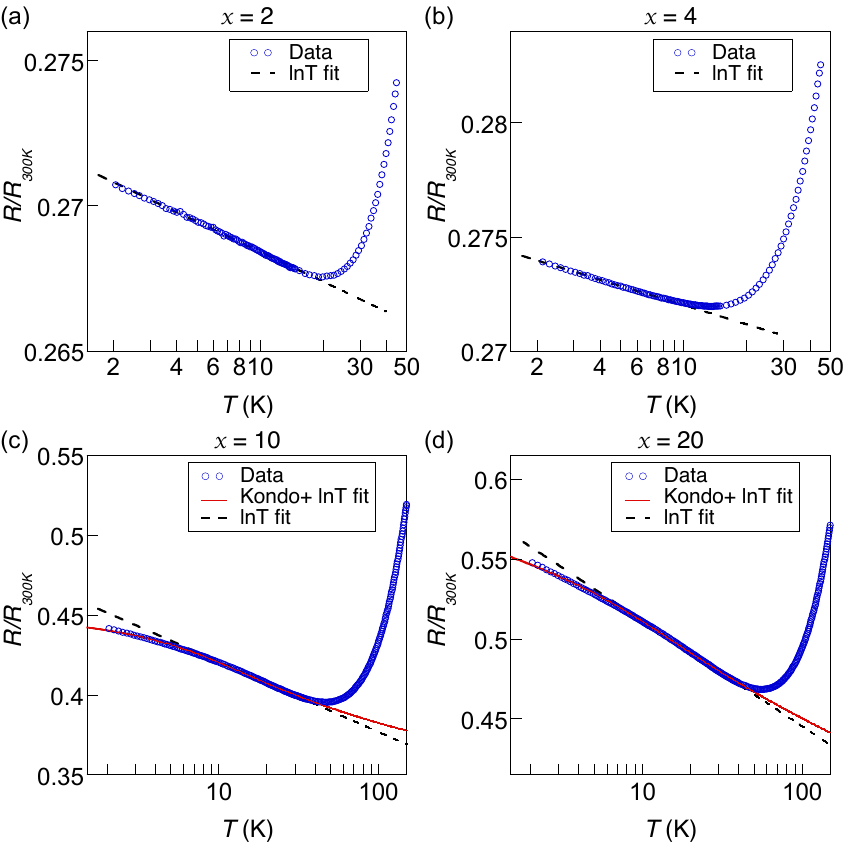}
\caption{\label{fig:RvT} Sample resistance $R$ (dots) as a function of temperature $T$ on logarithmic scale, normalized to its value at $T=300$~K, for the STO(8 u.c.)/NTO($x$ u.c.)/STO(8 u.c.)/LSAT(001) heterostructures, with $x=$2, 4, 10 and 20 respectively for panels  (a), (b), (c) and (d). Dashed lines show the low-temperature logarithmic dependencies and solid lines the theoretical fits using Eqs.~\ref{eq:RvT} and \ref{eq:R_K}.}
\end{figure}

Fig.~\ref{fig:RvT} shows the temperature dependence of resistance $R$, normalized to its value at $T=300$~K, for devices with four different thicknesses of the NTO layer. An upturn in resistance is observed at low temperatures for all the samples. For samples with 2 or 4~u.c.~NTO, the resistance upturn follows a logarithmic dependence on temperature down to 2~K (dashed lines in Figs.~\ref{fig:RvT}(a) and (b)), which is consistent with WAL/WL. In contrast, as the thickness of the NTO layer is increased to 10 and 20~u.c., the experimental resistance upturn deviates from and lies below the logarithmic dependence for $T<\sim$10~K (Figs.~\ref{fig:RvT}(c) and (d)). Saturation of the logarithmic dependence at low temperatures is characteristic of the Kondo effect, originating from the interplay between the conduction electrons and magnetic impurities.

We find that the $R$~vs.~$T$ behavior of 10 and 20~u.c.~samples at low temperatures can be well described by a simple model, where the Kondo contribution $R_K$ is taken into account, given as
\begin{equation}
\label{eq:RvT}
R(T)=R_0-q\ln T +R_K(T).
\end{equation}
The $\ln T$ term represents the non-saturating resistance upturn due to WAL/WL.
$R_0$ includes both a $T$-independent term from the WAL/WL contribution and the residual resistance due to sample disorder.
For the Kondo contribution, we adopt the empirical relation \cite{lee_electrolyte_2011,costi_transport_1994} 
\begin{equation}
\label{eq:R_K}
R_K(T)=R_K(T=0)\left[(T/T_K)^2(2^{1/s}-1)+1\right]^{-s},
\end{equation}
which is a universal function of $T$ in units of the Kondo temperature $T_K$. Here, $T_K$ is defined as the temperature at which the Kondo contribution to the resistance reaches half of its zero-temperature value $R_K(0)$. The value of $s$ depends on the spin of the impurity $S$ and we fix $s=0.225$ for $S=1/2$.
Eq.~\ref{eq:R_K} is known to work well for a broad range of temperatures, from the logarithmic-dependence dominated region at $T\ll T_K$ to the Fermi liquid region at $T\gg T_K$. The solid curves in Figs.~\ref{fig:RvT}(c) and (d) are the fitting results for our data using Eqs.~\ref{eq:RvT} and \ref{eq:R_K}. The extracted values of the fit parameters are listed in Table~\ref{tab:RvTfit}. We obtain a Kondo temperature $T_K \sim31$~K for the 10~u.c.~sample and $\sim33$~K for the 20~u.c.~sample. 

\begin{table}[t]
\caption{\label{tab:RvTfit}Parameters extracted from the fits using Eqs.~\ref{eq:RvT} and \ref{eq:R_K} to the measured $R$~vs.~$T$ curves for a typical 10 and 20~u.c. Sample.}
\begin{ruledtabular}
\begin{tabular}{ccccc}
$x$~(u.c.) & $T_K$~(K) & $R_K(T=0)$~($\Omega$) & $R_0$~($\Omega$) & $q$\\
\hline
10 & 30.7 & 96.4 & 683.4 & 9.07\\
20 & 32.7 & 112.4 & 1122.6 & 34.9\\
\end{tabular}
\end{ruledtabular}
\end{table}

\begin{figure*} 
\includegraphics{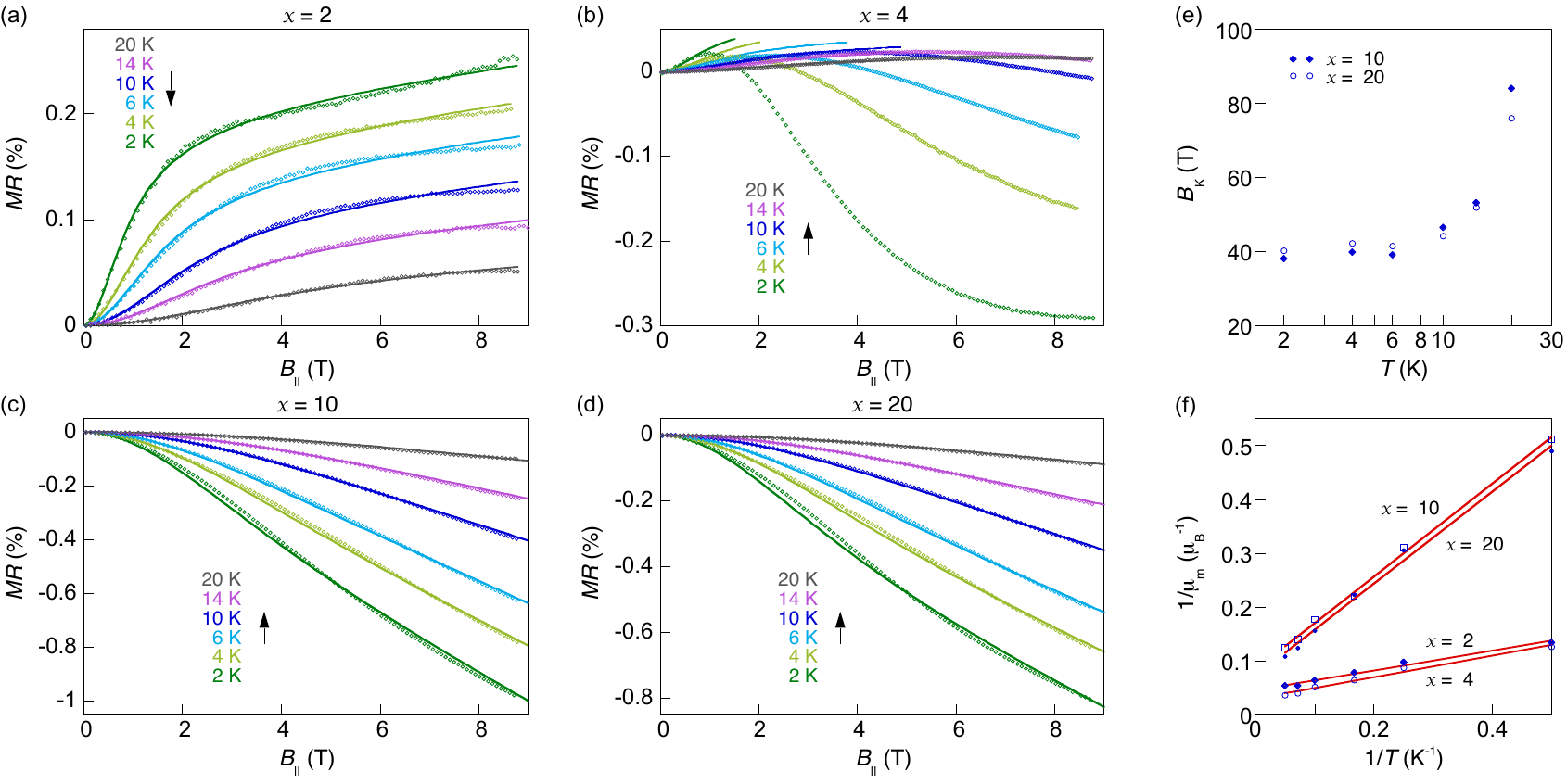}
\caption{\label{Fig3} Magnetoresistance as a function of the parallel field $B_{||}$ at various temperatures for samples with different NTO thicknesses, $x=$2 (a), 4 (b), 10 (c), 20~u.c.~(d): Experimental data (dots) and theoretical fits (solid lines). (e) The magnetic field scale $B_K$ of the Kondo contribution extracted from the fits in (c) and (d) for the 10 and 20~u.c. samples, respectively. (f) The inverse of the apparent local moment $1/\mu_m$ as function of $1/T$ (dots) and the associated linear fits (solid lines) for the samples in (a)-(d).}
\end{figure*}

The above analysis seems to suggest that both the localization and the Kondo effect contribute to the conductance of the interface in the thicker samples. To distinguish between the contributions, we measure the $MR$ in a magnetic field applied parallel to the sample plane for the low temperature range. In this case, the orbital effects are expected to be suppressed by the confinement in the $z$ direction due to the 2D nature of the system. This measurement therefore provides a platform for investigating the magnetic properties of the interfacial system.   
Figs.~\ref{Fig3}(a)-(d) show the experimentally measured $MR$ in a parallel magnetic field, $[R_s(B_{||})-R_s(0)]/R_s(0)\times 100\%$, for samples with different NTO thicknesses. As shown in Fig.~\ref{Fig3}(a), for a 2~u.c.~sample we observe pronounced positive $MR$ for the entire temperature range, with a sharp rise in resistance at low fields and a gradual increase at higher fields. As the NTO thickness increases to 4~u.c., while the low-field regime still features a positive $MR$, a high field regime characterized by a negative $MR$ appears for each temperature (Fig.~\ref{Fig3}(b)). The negative $MR$ regime is more noticeable at a lower temperature. As the NTO thickness increases further to 10 or 20~u.c., we find that the $MR$ remains negative over the entire field range (Figs.~\ref{Fig3}(c) and (d)). 

Previous analysis of the large positive $MR$ observed in 2~u.c.~samples revealed the significant roles played by the spin-orbit coupling and local magnetism on the conductance of the interface \cite{cai_disentangling_2019}. It is known that a parallel field can suppress WAL via the Zeeman interaction, resulting in positive $MR$. The presence of ferromagnetic order in localized regions at the interface gives rise to a substantial magnetic exchange interaction with the conduction electron spins. This exchange interaction effectively leads to a large Zeeman splitting and thus affects strongly the sample resistance. This phenomena can be assessed numerically based on localization theories \cite{cai_disentangling_2019}. The Zeeman interaction induces dephasing to the quantum interference in the presence of SOC, which can be described as a correction to $B_\phi$ \cite{minkov_weak_2004,malshukov_magnetoresistance_1997}, 
\begin{equation}
\label{eq:Zeeman}
\Delta_\phi(B_{||})=\frac{(g \mu_B B_{||})^2}{(4eD)^2B_{so}}.
\end{equation}
The interference-induced magnetoresistance in a parallel magnetic field is thus given by
\begin{equation}
\label{eq:ILP_inplane}
\frac{R_s(B_{||})-R_s(0)}{R_s(0)^2}= \frac{\sigma_0}{2}\ln\left[1+\frac{\Delta_\phi(B_{||})}{B_\phi}\right].
\end{equation}
The exchange interaction from local magnetic regions can be represented by an effective exchange field $B^E$, which couples to conduction electron spins, and enters the formulas via the Zeeman term \cite{dugaev_weak_2001}. As a result, the applied field $B_{||}$ in Eqs.~\ref{eq:Zeeman} and \ref{eq:ILP_inplane} is replaced by a total effective field in the plane, $B^t_{||}=B_{||}+B^E_{||}$. In a superparamagnetic system, the fluctuating moments $\mu_m$ of local magnetic regions tend to align along the applied magnetic field, leading to a net magnetization \cite{ohandley_modern_2000}. In this case, the applied field dependence of the exchange field is given by the Langevin function $L(x)$ at temperatures above the blocking temperatures, yielding $B^E_{||}(B_{||})=\lambda\mu_0 M_s L(\mu_m B_{||}/k_B T)$, where $\lambda$ is the coefficient characterizing the effective exchange interaction between electrons and the local moments, $\mu_m$ is the moment of a single magnetic region, and $M_s$ is the saturation magnetization. Using $\lambda \mu_0 M_s$ and $\mu_m$ as two variables, the  $MR$ data for the 2~u.c.~sample are well reproduced by the fits incorporating the Langevin function into Eqs.~\ref{eq:Zeeman} and \ref{eq:ILP_inplane} as described above. The fitting results are shown as solid lines in Fig.~\ref{Fig3}(a) for various temperatures. Here, $g m^*/m_e=1.5$ is assumed for the theoretical fits \cite{cai_disentangling_2019}. The values of $B_{so}$ and $B_{\phi}$ are obtained from the fitting of the $MR$~vs.~$B_{\bot}$ data for the same sample. We note the above analysis is limited to the condition $\Delta_\phi<B_{so}$, according to the localization theory \cite{malshukov_magnetoresistance_1997}. This condition is found to hold for our data in the entire measurement range presented in Fig.~\ref{Fig3}(a).

As the NTO thickness is increased, a negative component in the $MR$ emerges and eventually dominates the sample behavior in the parallel magnetic field (Figs.~\ref{Fig3}(b)-(d)). 
The large negative $MR$ observed under $B_{||}$ for the 10 and 20~u.c.~samples cannot be explained by the WAL/WL effect (see the supplementary material for details). 
It is well known that the Kondo effect leads to negative $MR$. The specific expression for the Kondo resistance as a function of the applied magnetic field, $R_K(B/B_K)$, has been derived by the Bethe-ansatz approach for zero temperature and is a universal function of $B$ in units of a magnetic scale $B_K$  \cite{andrei_solution_1983,lee_electrolyte_2011}. The two distinct scales $T_K$ and $B_K$ are related. The ratio of $k_B T_K$ to $\mu_B B_K$ was found to be of order one \cite{furuya_bethe-ansatz_1982}.  
Further assuming that the Kondo contribution to sample resistance dominates the $MR$ behavior, one can obtain the expression for the case of parallel magnetic field $B_{||}$,
\begin{equation}
\label{eq:MR_inplane}
\frac{R(B_{||})-R(0)}{R(0)}= - \frac{R_K(0)}{R(0)}\left(1-\cos^2\left[\frac{\pi}{2}M^i(B_{||}/B_K)\right]\right),
\end{equation}
where $R(0)$ and $R_K(0)$ are the zero-field value of the total sample resistance and that of the Kondo contribution, respectively (see the supplementary material for the exact form of the function $M^i$). The negative $MR$ changes quadratically with $B_{||}$ for $B_{||}\ll B_K$ and decreases more gradually for $B_{||}\gg B_K$. 

To incorporate the effect of local magnetism into the Kondo model, we adopt a similar approach as described above by introducing the exchange field $B^E_{||}$. As a result, the applied field $B_{||}$ in Eq.~\ref{eq:MR_inplane} is replaced by a total effective field in the plane, $B_{||}+B^E_{||}$, where the $B_{||}$ dependence of $B^E_{||}$ is given by the Langevin function. Using $B_K$, $\lambda \mu_0 M_s$ and $\mu_m$ as three fit variables, the negative $MR$ observed in the 10 and 20~u.c.~samples are well reproduced by this model. The fitting results are shown in Figs.~3(c) and (d) for various temperatures. We should note that here, the zero-field value of $R_K(0)/R(0)$ in Eq.~\ref{eq:MR_inplane} is obtained for each temperature from the $R$~vs.~$T$ fit of the same sample (Figs.~\ref{fig:RvT}(c) and (d)) and is not a free parameter. 

Fig.~\ref{Fig3}(e) shows the extracted values of the Kondo magnetic scale $B_K$ as a function of temperature for the 10 and 20~u.c. samples.
At temperatures above $\sim10$~K, $B_K$ decreases monotonically with decreasing temperature. As the temperature is further reduced below $\sim10$~K, $B_K$ approaches its true value and becomes stabilized around 40~T. This trend is expected as Eq.~\ref{eq:MR_inplane} works explicitly for temperatures significantly smaller than where the Kondo impurity $S=1/2$ is mostly screened.
Moreover, we note that the order of magnitude for the extracted magnetic energy scale matches that of the Kondo temperature extracted from $R$~vs.$T$ data. The ratio between the two is $(k_B T_K)/(\mu_B B_K)\approx 1.2$ for our systems. 

The 4 u.c.~sample is in the crossover regime where both positive and negative $MR$ can be seen (Fig.~\ref{Fig3}(b)). For simplicity, we fit only the positive $MR$ at low fields where the WAL contribution dominates, using the fit method described above for the 2~u.c.~sample. The fitting results are presented as the solid lines in Fig.~\ref{Fig3}(b).

The above analyses on the magneotresistance under parallel field provide a quantitative assessment of the magnetic moments of individual local magnetic regions at the interfaces. The extracted values of $\mu_m$ for all the four samples are shown in Fig.~\ref{Fig3}(g). The inverse of the apparent moment $1/\mu_m$ changes linearly with the inverse of the temperature. This behavior is consistent with the scenario of weakly interacting superparamagnets. The true moment $\mu_m^*$ is given by the relation $1/\mu_m=1/\mu^*(1+T^*/T)$, where $T^*$ charaterizes the energy scale of the dipole-diple interaction \cite{allia_granular_2001}. From the linear fits in Fig.~\ref{Fig3}(g), we obtain the average magnetic moment of an individual nanoscale magnetic region to be $\mu_m^*=$22, 33, 12 and 14$\mu_B$ for the 2, 4, 10 and 20~u.c.~samples, respectively. 


In summary, we investigate the low-temperature magneto-transport in a series of our hetero-interfaces with different thicknesses of the NTO layer. We provide a consistent analysis on the measurement results, including the temperature dependence of resistance, magnetoresistance in a perpendicular magnetic field, and the magnetoresistance in a parallel field, for all the samples (the data and analysis for the magnetoresistance in the perpendicular field can be found in the supplementary material). Our analysis shows that increasing the thickness of the NTO layer introduces substantial Kondo scattering at the interface, which dominates over WAL in the thickest samples studied. Intriguingly, we find that the effects of the SOC and the Kondo scattering on the conduction electrons are altered and enhanced by the presence of the local magnetic order, leading to distinct magneto-resistance behaviors in a parallel magnetic field that serve as a sensitive probe of the local magnetic landscape. The co-existence of the Kondo effect and magnetic exhange fields also identifies NTO/STO interfaces a promising system for exploring the interplay between atomic-scale spin defects, nanoscale magnetically-ordered regions and spin-orbital coupling in two dimensions.     

\

This work was supported primarily by the Office of Naval Research under Award No. N00014-17-1-2884. Film growth and structural characterizations were funded by the U.S. Department of Energy through the University of Minnesota Center for Quantum Materials, under Grant No. DE-SC-0016371. Portions of this work were conducted  in the Minnesota Nano Center, which is supported by the National Science Foundation through the National Nano Coordinated Infrastructure Network (NNCI) under Award Number ECCS-1542202. Sample structural characterization was carried out at the University of Minnesota Characterization Facility, which receives partial support from NSF through the MRSEC program under Award No. DMR-1420013. The authors would like to thank Zhen Jiang and Xiaojun Fu for assistance with PPMS measurements and thank Yilikal Ayino, Xuzhe Ying, Amanda G. Lamas, and Paul Crowell for valuable discussions. 

\bibliography{Refs}

\end{document}